\documentclass{pasj00}
\draft

\def\ltsima{$\; \buildrel < \over \sim \;$}
\def\simlt{\lower.5ex\hbox{\ltsima}}            
\def\gtsima{$\; \buildrel > \over \sim \;$}
\def\simgt{\lower.5ex\hbox{\gtsima}}

\newcommand{\erg}{erg~s$^{-1}$}
\newcommand{\ergs}{erg~cm$^{-2}$~s$^{-1}$ }

\begin{document}
\SetRunningHead{Author(s) in page-head}{Running Head}
\Received{2002/06/25}
\Accepted{2002/08/12}

\title{Chandra Observations and Optical Identification of Hard
X-ray Sources Discovered with ASCA}

\author{
Shin \textsc{Watanabe}\altaffilmark{1,2}
Masayuki \textsc{Akiyama}\altaffilmark{3}
Yoshihiro \textsc{Ueda}\altaffilmark{1} \\
Kouji \textsc{Ohta}\altaffilmark{4}
Richard \textsc{Mushotzky}\altaffilmark{5}
Tadayuki \textsc{Takahashi}\altaffilmark{1,2}
and
Toru \textsc{Yamada}\altaffilmark{6}
}

\altaffiltext{1}{Institute Space and Astronautical Science, Sagamihara, 
Kanagawa, 229-8510}
\altaffiltext{2}{Department of Physics, University of Tokyo, Tokyo, 113-0033}
\altaffiltext{3}{Subaru Telescope, National Astronomical Observatory of Japan,
Hilo, HI, 96720, USA}
\altaffiltext{4}{Department of Astronomy, Kyoto University, Kyoto, 606-8502}
\altaffiltext{5}{NASA Goddard Space Flight Center, Greenbelt, 
Maryland, 20771, USA}
\altaffiltext{6}{National Astronomical Observatory of Japan, Mitaka, 
Tokyo, 181-8588}

\email{watanabe@astro.isas.ac.jp}


\maketitle

\begin{abstract}

We present the first results of the Chandra and optical follow-up
observations of hard X-ray sources detected in the ASCA Medium
Sensitivity Survey (AMSS). Optical identifications are made for five
objects. Three of them show either weak or absent optical narrow
emission lines and are at low redshift $<$z$>$~$\sim$~0.06. One of
them is a broad line object at z~=~0.910 and one is a z~=~0.460 object
with only narrow lines. All the narrow line objects show strong
evidence for absorption in their X-ray spectra. Their line ratios are
consistent with a Seyfert II/LINER identification as are the line
widths. The three low redshift objects have the colors of normal
galaxies and apparently the light is dominated by stars. This could be
due to the extinction of the underlying nuclear continuum by the same
matter that absorbs X-rays and/or due to the dilution of the central
source by starlight. These results suggest that X-ray sources that
appear as ``normal'' galaxies in optical and near-IR bands
significantly contribute to the hard X-ray background. This population
of objects has a high space density and probably dominates the entire
population of active galaxies.

\end{abstract}

\section{Introduction}

The origin of the Cosmic X-ray Background (CXB) has been one of great
mysteries in X-ray astronomy. At soft X-ray energies (below 2~keV),
the ROSAT surveys resolved 70--80~\% of the CXB into discrete sources
(e.g., \cite{Hasinger1998}). The majority of the optical
identifications are unobscured active galactic nuclei (AGNs), mostly
Seyfert Is and quasars (\cite{Lehmann2001}). At higher energies, an
additional population of absorbed or flat spectrum objects are needed
to make up the CXB, which has a flatter spectrum in the 2--10~keV band
than the unobscured AGNs.

Hard X-ray surveys performed with ASCA resolved about 30~\% of the CXB
in the 2--10~keV band, revealing the presence of a population of hard
sources emerging at flux levels of several 10$^{-13}$~\ergs\
(\cite{Ueda1999a}; \cite{Ueda1999b}). Recently, almost all of the hard
X-ray background has been resolved into point sources by the
Chandra deep surveys (e.g.,
\cite{Mushotzky2000}; \cite{Brandt2000}; \cite{Giacconi2001}). The
faint sources detected with Chandra have, on average, rather hard
spectra (e.g., \cite{Rosati2002}), which indeed account for the
spectrum of the CXB. However, due to the faintness of the Chandra
sources in the optical band, their nature still remains, in general
unclear (\cite{Barger2001}).

In order to understand this hard X-ray population over a broad range
in  luminosity and
redshift , it is necessary to reveal the nature of the
``bright'' sources (i.e., with fluxes above $10^{-13}$~\ergs\ in the
2--10~keV band) that make up about 30~\% of the CXB. In addition,
because of their brightness, these sources should be much easier to
study over a  wide range of wavelengths than the ultra-faint Chandra sources, 
and serve as  a
prototype of the key populations that comprise the CXB. The results of
the optical identification of a complete hard-band selected sample in the
ASCA Large Sky Survey (ALSS, \cite{Akiyama2000}) indicates that the
majority of the hard sources are nearby $(z\simlt 0.5)$ type-2
AGNs. However, in particular when sources with hard X-ray spectra are
concerned, our knowledge is still quite limited due to the small sample
size.

Thus, we have started to study the optical counterparts of the
``bright'' and ``hard'' sources selected from the ASCA Medium
Sensitivity Survey (AMSS; \cite{Ueda2001}), a serendipitous source
survey of the extragalactic sky based on the ASCA GIS archival data at
flux limits of 10$^{-13}$~\ergs\ (2--10~keV). The advantage of the
AMSS is its large solid angle of sky coverage, which is indispensable
for obtaining a reasonable sample of ``bright'' (hence rare)
sources. 

In this paper we report the first results of the Chandra observations
and optical identifications for hard AMSS sources selected in Chandra
AO-1.
The excellent positional accuracy of
$\sim$~1~arcsec obtained with Chandra has enabled us to make
unambiguous optical identifications of these sources. We present the results 
from X-ray,
optical, and near infrared observations, in \S~2, 3 and 4,
respectively, followed by detailed description of individual sources
in \S~5. We discuss the results in \S~5 and our conclusions are given in \S~6. 
Throughout the paper, we assume a Hubble constant of 
$H_0$~=~50~km~s$^{-1}$~Mpc$^{-1}$ 
and a deceleration parameter of $q_0$~=~0.5. 
Optical identification for a complete flux-limited sample using
a part of the AMSS catalog is reported by \citet{Akiyama2002}.

\section{X-ray Data}

\subsection{The Sample}

Table~\ref{list_tab} lists the seven sources that were selected for
Chandra AO1 observations.
We selected them from an AMSS catalog with
the following two criteria\footnote{In the subsequent Chandra AO2 and
AO3 observations we select targets with slightly different criteria
and strategy from the published AMSS catalog, both aiming at somewhat
brighter sources with fluxes above $3~\times~10^{-13}$~\ergs\ (2--10~keV). 
The details will be given in future papers.}
:
(1) Objects are detected in the 2--10~keV band with a flux
fainter than $3\times 10^{-13}$~\ergs\ (assuming a photon index of 1.7), but (2) not detected in the 0.7--2~keV band, with detection significance above
4.5~$\sigma$.
None of the sources have ROSAT HRI or PSPC counterparts when the ROSAT
image is available, and the best-fit photon index in the ASCA data is
less than 0.8, consistent with or flatter than the X-ray background
spectral slope within the error. 
Because this sample is
selected essentially without any bias, they are considered to be a
good representative of the X-ray population with hard X-ray spectra at
fluxes of (1--3)~$\times~10^{-13}$~\ergs\ (2--10~keV). 
We selected these sources from an
earlier version of the AMSS catalog than the published one
(\cite{Ueda2001}). Also, we adopted a slightly lower threshold for
detection significance (4.5~$\sigma$) than that used for the published
catalog (5.0~$\sigma$). As a result, some sources in this sample do not
exist in the catalog published by \citet{Ueda2001}. The corresponding
source names of the AMSS catalog (with a prefix of ``1AXG''), if
present, are listed in Table~\ref{list_tab}. 
None of them are
in the sample of \citet{Akiyama2002}, who use slightly brighter
sources with more conservative detection criteria. The total number of
2--10 keV~detected ($\geq 5\sigma$) sources at Galactic latitudes $|b|
> 20^\circ$ that have similar properties (i.e., with the apparent
photon index smaller than 0.8 at the flux range of 
(1--3)~$\times~10^{-13}$~\ergs ) is 
18 out of about 800 serendipitous sources in the
\citet{Ueda2001} catalog.

\subsection{Chandra Results}

The log of Chandra observations is given in Table~\ref{list_tab}. All
the observations were performed with the ACIS detector
(\cite{Bautz1998}), placing the target near the aim point of the
ACIS-I array on the CCD~I3 chip. We analyzed the data in a standard
manner using the {\it ``CIAO 2.1''} software. We performed source
search by the {\it ``celldetect''} program, which uses a simple
sliding cell method. From all the seven fields, we detected X-ray
sources in the ASCA error circle. We found, however, that the results
for two of the targets are rather different than expected from the
ASCA results. The Chandra 2--10 keV flux of AXJ~1227+4421
(10$^{-14}$~\ergs) is more than ten times less than the ASCA
flux. Similarly, in the error circle of AXJ~2018+1139, two X-ray
sources were detected with fluxes of 10$^{-14}$~\ergs , whose summed
flux does not account for the ASCA-measured flux. Possible reasons for
the discrepancy are discussed in \S~5.1. In this paper, we concentrate
on the other five sources that have clear Chandra counterparts for
which optical identification was made.

Table~\ref{xray_property_tab} gives the X-ray properties of the five
targets. The positions are accurate within $<$~1~\arcsec 
\footnote{Chandra science center memo
http://cxc.harvard.edu/mta/ASPECT/celmon/index\_update.html}. 
For reference we list ``apparent'' power-law photon indices from the
hardness ratio $(H-S)/(H+S)$, where $H$ and $S$ is the count rate in
the 2--10~keV and 0.7--2~keV bands, respectively, correcting for the
Galactic absorption. 

\begin{figure}
 \begin{center}
  \FigureFile(70mm,60mm){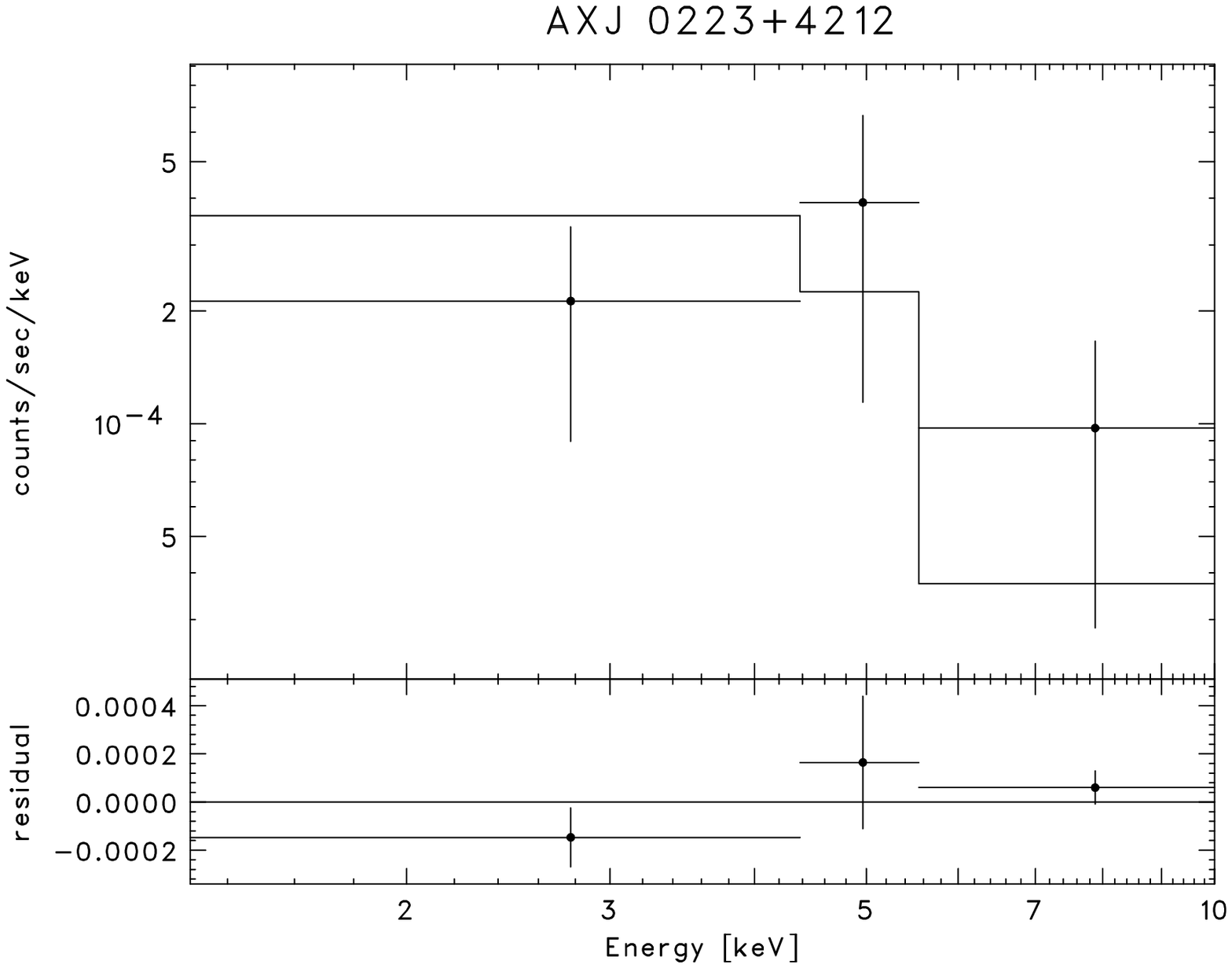}
  \FigureFile(70mm,60mm){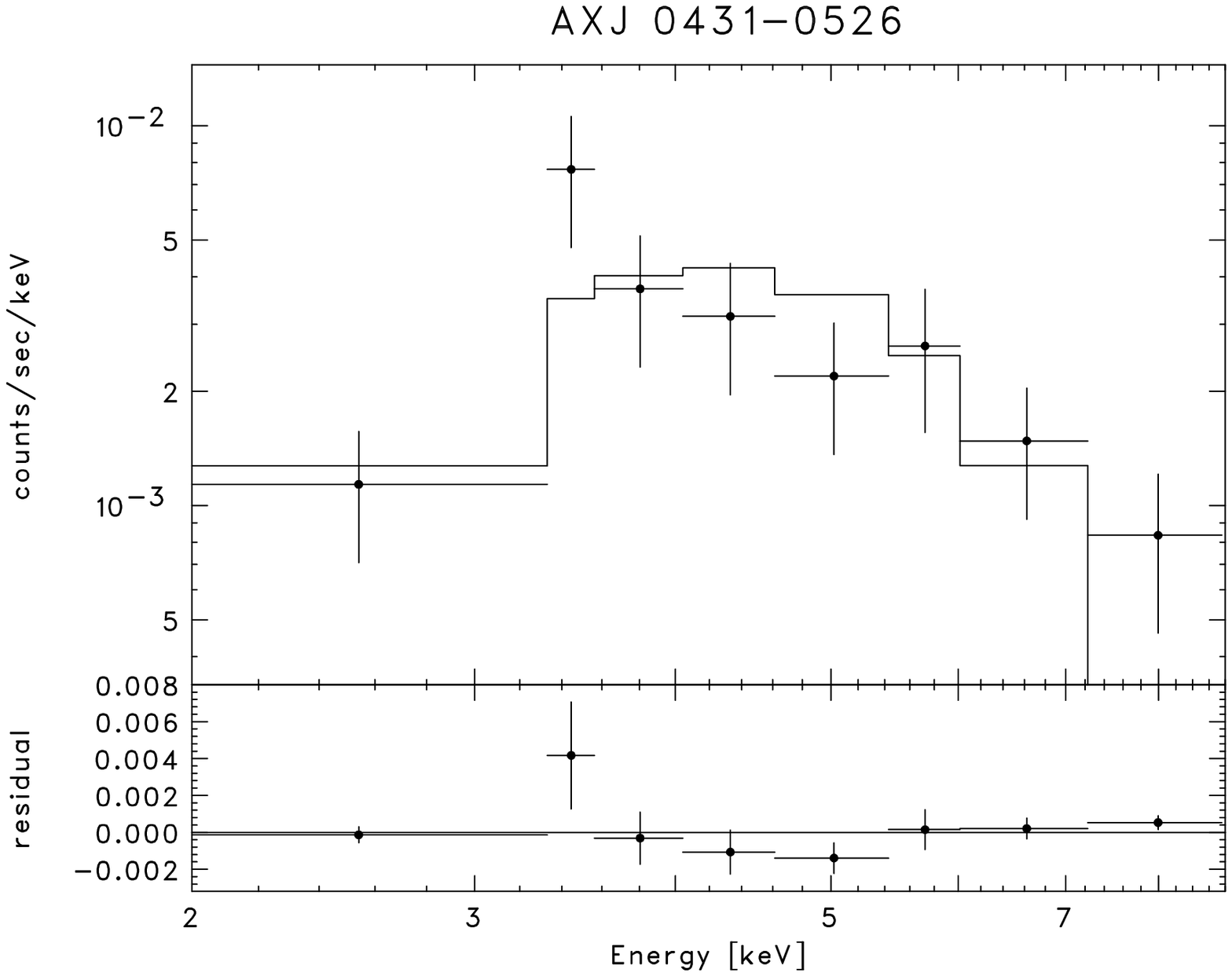}

  \FigureFile(70mm,60mm){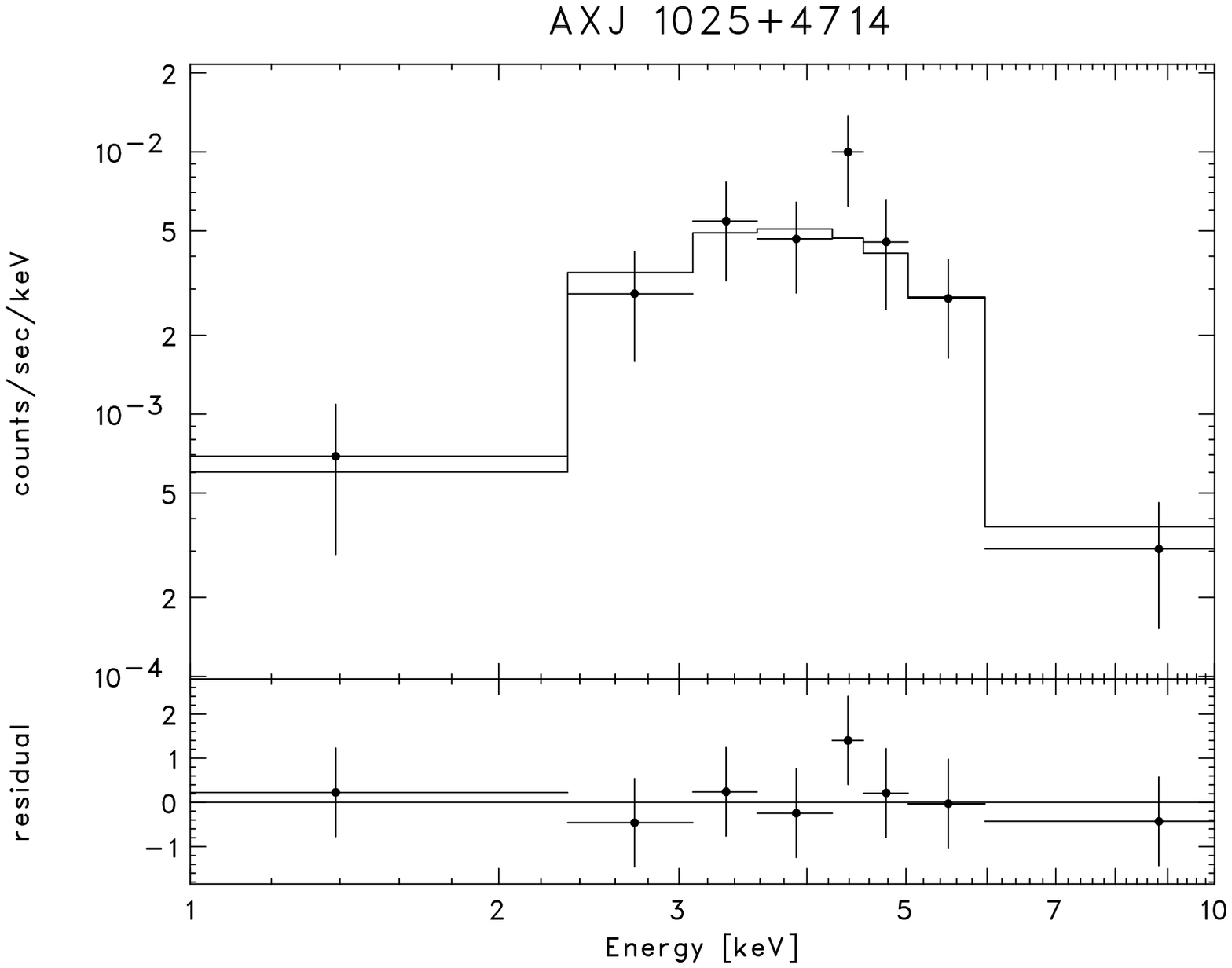}
  \FigureFile(70mm,60mm){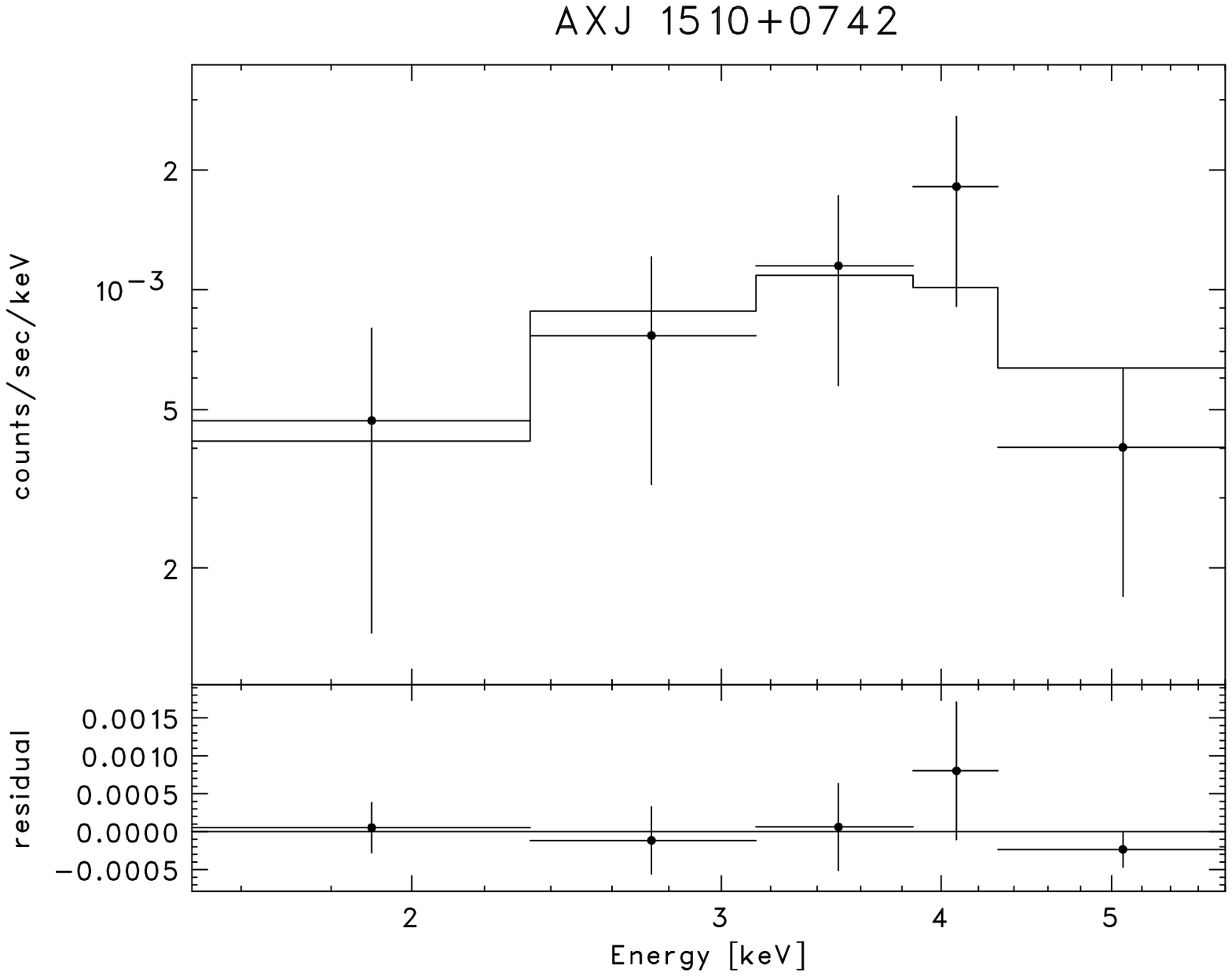}

  \FigureFile(70mm,60mm){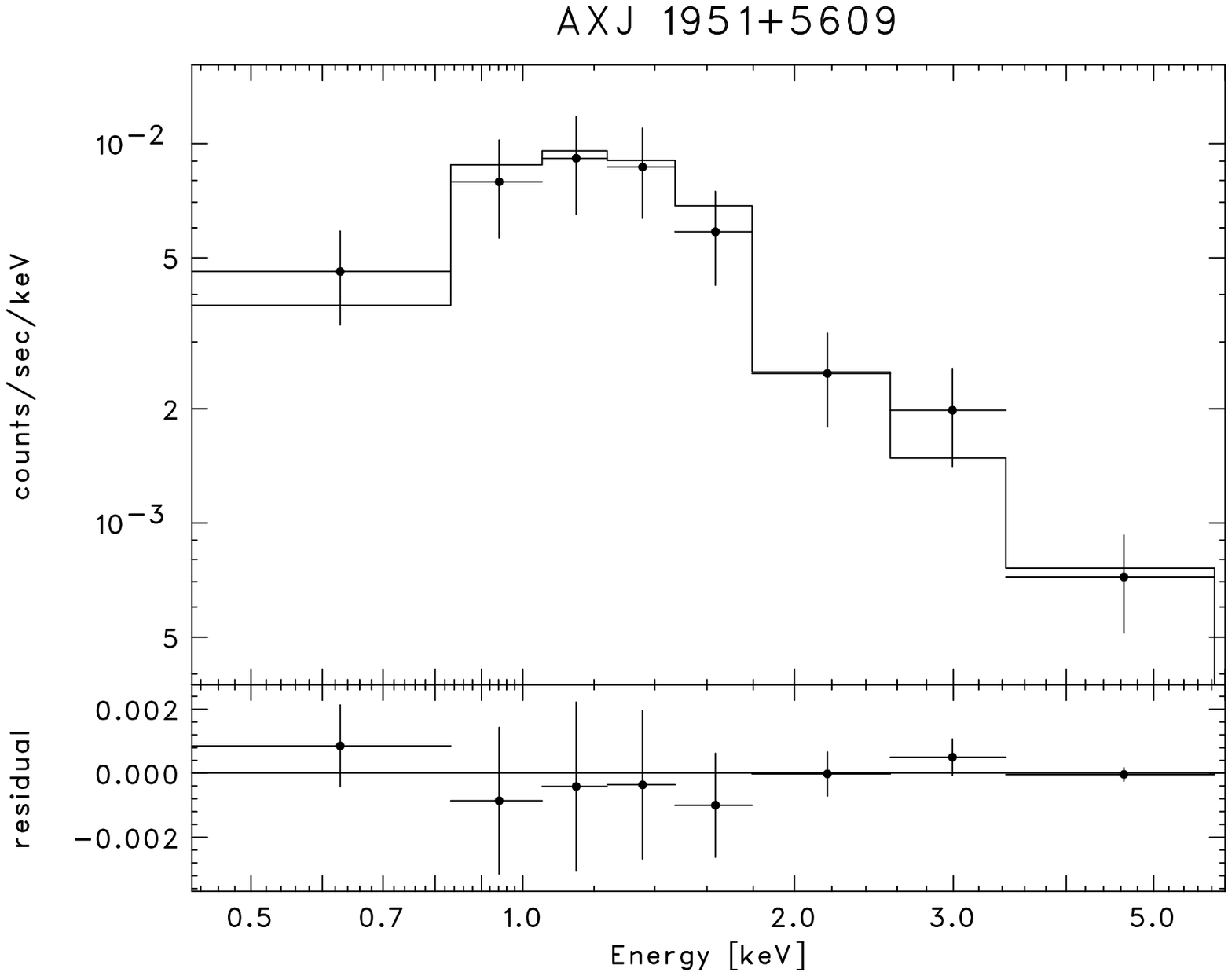}

\caption{The ACIS spectra of AXJ~0223$+$4212, AXJ~0431$-$0526, AXJ~1025+4714,
AXJ~1510+0742, and AXJ~1951+5609, folded with the response. 
The best-fit model is an absorbed power law with a photon index of 1.7.
}
\label{xray_spec_fig} \end{center}
\end{figure}

Fig.~\ref{xray_spec_fig} shows the ACIS pulse height spectra.  Using
the redshift determined from the optical identification (see \S~3), we
calculated the amounts of absorption at the source redshift assuming a
photon index fixed at 1.7, a typical value for AGNs (e.g.,
\cite{Mushotzky1993}). Because the background is negligible and the number
of photons in each bin is very small, we used the Poisson likelihood
statistics to search for the best-fit parameters and the errors. We
find that the column densities for AXJ~0223+421, AXJ~0431$-$0526,
AXJ~1025+4714 and AXJ~1510+0742 are $N_{\rm H} \sim
10^{22-23}$~cm$^{-2}$, and that for AXJ~1951+5609 is
$\sim$~10$^{21}$~cm$^{-2}$.  Using the best-fit parameters, we
calculate the observed 2--10 keV flux and the absorption-corrected
2--10 keV luminosity. These results are also summarized in
Table~\ref{xray_property_tab}. The range of X-ray luminosity is
$\sim$10$^{41-45}$ erg~s$^{-1}$.

\begin{figure}
 \begin{center}
  \FigureFile(80mm,60mm){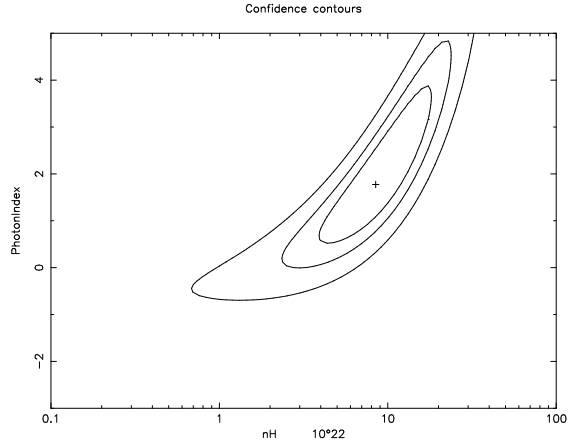}
  \caption{The confidence contour of the photon index and the column
density at the source redshift constrained from the ACIS spectrum of 
AXJ~1025+4714. Each contour represents 68 \%, 90 \% and 99 \% confidence level.} \label{cont_fig}
 \end{center}
\end{figure}

Since AXJ~1025+4714 has sufficient photon counts, we fit its spectrum
with both the photon index and column density being free
parameters. Fig.~\ref{cont_fig} shows the confidence contour for the
two parameters. This indicates that the hard spectrum cannot be
attributed purely to an intrinsically flat spectrum and a significant
amount of absorption is indeed required.

\section{Optical Data}
\subsection{Observations}

We conducted optical follow-up observations for the five Chandra
sources corresponding to AXJ~0223$+$4212, AXJ~0431$-$0526,
AXJ~1025$+$4714, AXJ~1510$+$0742, and AXJ~1951$+$5609. Firstly,
optical counterparts of the Chandra sources are selected from the
Digitized Palomar Observatory Sky Survey plate (hereafter DPOSS). All
the five sources have only one optical counterpart above the limit of
the DPOSS. Based on the range of optical to X-ray fluxes one
anticipates that, as opposed to the fainter Chandra and XMM sources,
the optical counterparts of the ASCA sources are expected to be in the
15--18th mag range, well above the flux limits of the DPOSS (see
Figure~3 of \citet{Hasinger1999}, which shows the observed range of
X-ray to optical fluxes.) and thus their detection in the DPOSS images
is expected.
Spectroscopic observations of the optical counterparts were made
during the course of optical follow-up observations for a complete
hard-band selected sample of the AMSS (\cite{Akiyama2002}). A detailed
journal of observation of each source is shown in
Table~\ref{optjarnal_tab}.

\begin{figure}
\begin{center}

\FigureFile(40mm,40mm){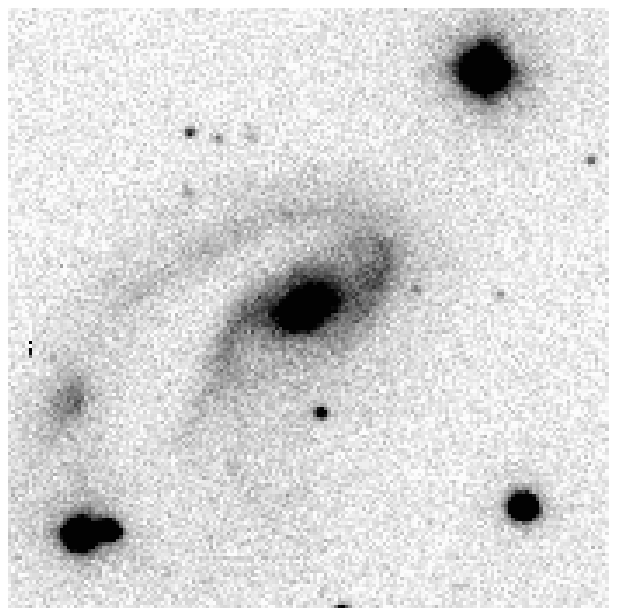}
\FigureFile(40mm,40mm){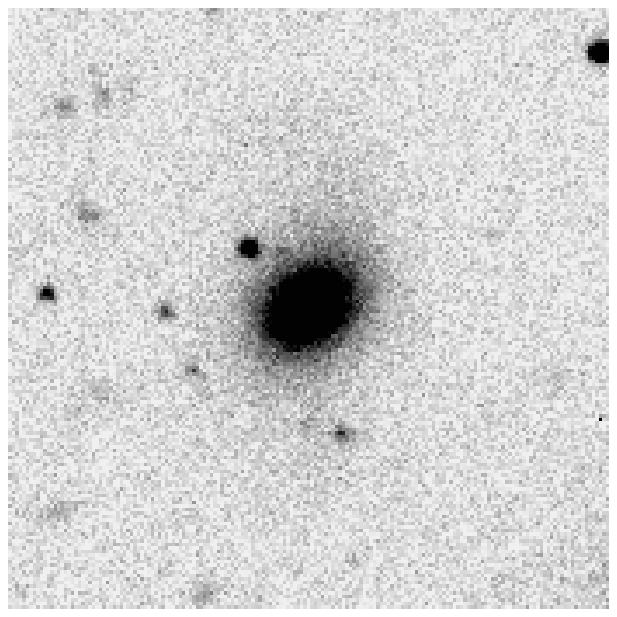}
\FigureFile(40mm,40mm){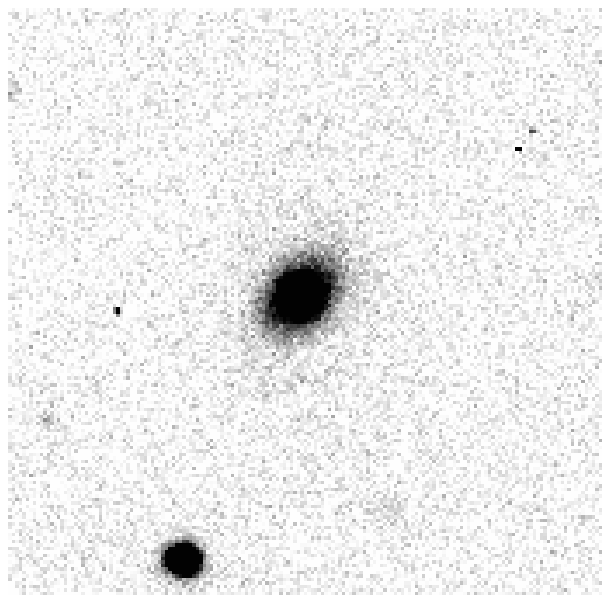}
\FigureFile(40mm,40mm){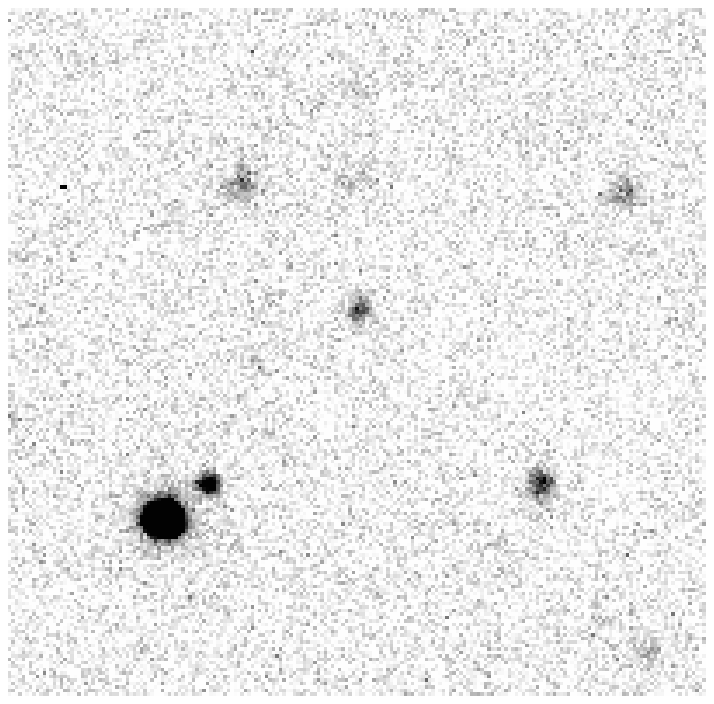}

\caption{No-filter 60~\arcsec~$\times$~60~\arcsec images of
AXJ~0223+4212, AXJ~0431$-$0526, AXJ~1025+4714, and AXJ~1951+5609 (from
left to right).}
\label{opt_image_fig}

\end{center}
\end{figure}

For AXJ~0223$+$4212, AXJ~0431$-$0526, AXJ~1025$+$4714, and
AXJ~1510$+$0742, spectroscopic observations with the University
of Hawaii 88$^{\prime\prime}$ telescope were obtained on 5, 6 
October 2000, and 20, 21 March 2001, 
respectively. We used the Wide Field Grism Spectrograph
with a grating of 420 grooves mm$^{-1}$ and the blaze wavelength of
6400~\AA. The spatial resolution was 0.\arcsec 35~pixel$^{-1}$ and the
typical image size during the observations was 0.\arcsec 8 --
1.\arcsec 2. A slit width of 1.\arcsec 2 was used. We covered the
wavelength range from 4000~\AA~ to 8000~\AA~ without an order cut
filter. The spectral resolution, which was measured by the HgAr lines
in comparison frames and night-sky lines in object frames, was 12~\AA~
(FWHM) which corresponds to 500 km~s$^{-1}$ at 7000~\AA. We took
imaging data of each object without any filter with higher spatial
resolution than the DPOSS. Fig.~\ref{opt_image_fig} shows the no-filter
60\arcsec~$\times$~60\arcsec~ images of the four sources.

For AXJ~1951$+$5609, the spectroscopic observation was done at the
KPNO 2.1~m telescope on 20 September 2000.  We used the Gold Camera
Spectrograph with Grating (\#32) with 300 grooves mm$^{-1}$ and blaze
wavelength of 6750~\AA.  The spatial sampling was 0.\arcsec 78~pixel$^{-1}$.
The typical image size during the observation was 
2.\arcsec 0. The slit width was 2.\arcsec 0 for objects and
10~\arcsec~ for standard star observations. The dispersion was
2.47~\AA~pixel$^{-1}$. We covered the wavelength range from 4000~\AA~
to 8000~\AA~ with an order cut filter for the wavelength range shorter
than 4000~\AA~ (GG400).  The spectral resolution was measured to be
8~\AA~ (FWHM) from night-sky lines in object frames.

\subsection{Analysis and Results}

All of the data were analyzed using IRAF\footnote{IRAF is distributed
by the National Optical Astronomy Observatories, which is operated by
the Association of Universities for Research in Astronomy, Inc. (AURA)
under cooperative agreement with the National Science
Foundation.}. After bias subtraction, flat-fielding, and wavelength
calibration, optimum extraction method with the {\it ''apextract''}
package was used to extract one dimensional spectral data from the two
dimensional original data. Flux calibration of the one dimensional
spectral data was made with spectra of standard stars (Feige~34 and
BD+28 are standard stars for March and October runs, respectively).
The finding chart made from the DPOSS and the flux calibrated optical
spectrum of each object are shown in Fig.~\ref{finding_fig}. Basic
parameters of the sample are listed in Table~\ref{basic_tab}.

Based on flux-calibrated spectrum of each object, we estimated the
magnitude of the objects. The flux density at 6400~\AA~ of each object
is converted to the $R$--band magnitude. The estimated $R$--band
magnitudes are listed in Table~\ref{basic_tab}.
Because the slit aperture of the spectroscopic observation does 
not cover the whole area 
of the objects, the magnitudes represent those in that region.
The uncertainty of the estimation of the magnitude is
$\pm$~0.9 magnitude without the slit-aperture loss.

While all of the optical counterparts show emission lines, only
AXJ~1951$+$5609 shows a strong broad emission line. To evaluate the
strength and the velocity width of the narrow emission lines of the
other four sources quantitatively, we performed spectral fitting for
the lines using a $\chi^{2}$ minimization method with the {\it
``specfit''} command in the {\it ``spfitpkg''} package of IRAF. The
FWHMs of line widths were deconvolved by the spectral resolution
mentioned above and are given in Table~\ref{line_equi}.  Measured flux
ratios of the [OIII]$\lambda$ 5007 to H$\beta$ the [NII]$\lambda$ 6583
to H$\alpha$, and the [SII]$\lambda\lambda$ 6717, 6731 to H$\alpha$
are also listed in Table~\ref{line_equi}, and are plotted in
Fig.~\ref{line_ratio_fig}. Solid lines in the figure represent
boundaries between Seyfert II galaxies, LINERs, and H~II region-like
galaxies taken from \cite{Veilleux_and_Osterbrock1987} and
\cite{Veilleux1995}.

\section{Near-Infrared Data}

AXJ~0223$+$4212, AXJ~0431$-$0526, and AXJ~1025$+$4714 are detected in
the 2MASS survey \footnote{http://www.ipac.caltech.edu/2mass/} 
and their $J$, $H$, and $K_s$ band
magnitudes are summarized in Table~\ref{2mass_tab}.
For AXJ~1510$+$0742 and AXJ~1951$+$5609, we conducted near-infrared
photometric observation with the SQIID (Simultaneous Quad Infrared
Imaging Device) infrared camera with four 512$\times$512 ALADDIN InSb
arrays attached to Kitt Peak National Observatory 2.1~m telescope on
31 March 2002. $J$--, $H$--, $K_s$--, and $L$--band images were taken
simultaneously but here we use only $J$--, $H$--, and $K_s$--band data.
The pixel scale of the camera was 0.$^{\prime\prime}$69~pixel$^{-1}$.
The FWHM of point sources 
during the observation was 1.$^{\prime\prime}$3,
1.$^{\prime\prime}$4, and 1.$^{\prime\prime}$5, in the $J$--, $H$--, and
$K_s$--bands, respectively. 
In each band, five frames each of which consists of 18 (for
AXJ~1510$+$0742) or 12 (for AXJ~1951$+$5609) coaddition of 10s
integration were taken, thus in total, the effective integrations were
900~s and 600~s for AXJ~1510+0742 and AXJ~1951+5609, respectively.
The pixel scale of the camera was 0.$^{\prime\prime}$69~pixel$^{-1}$.
The seeing condition during the observation was 1.$^{\prime\prime}$3,
1.$^{\prime\prime}$4, and 1.$^{\prime\prime}$5, in the $J$--, $H$--, and
$K_s$--bands. 
During the observation, UKIRT faint standard stars (FS11,
FS16, FS17, FS21, FS23, FS27) were observed for photometric
calibration. For each star, we took 5 images with 5 coaddition of 10~s
integration, changing the position of the star on the detector to
reduce any systematic errors.

The data reduction were done in the following steps. At first, we
subtracted a dark frame, which is made from a stacking of 10 images of
a cold shutter with the same integration time, from the object
data. The dark-subtracted frames were flat-fielded with a flat image,
which is made by stacking the object frames with normalizing their
background level. Finally, after correcting the offset of each object
frame, we combined 5 frames of the each object. Dark-subtraction and
flat-fielding were done for standard star frames in the same manner,
but we did not shift and combine the standard star frames. Counts of
the standard star in each frame were measured individually by using a
growth curve fitting method for each star. Based on the scatter of the count
rate to magnitude conversion factors derived from the observed UKIRT
faint standard stars, we estimated the uncertainties of the
photometric calibration as 0.02~mag, 0.04~mag, and 0.02~mag in $J$--,
$H$--, and $K_s$--bands, respectively. The measured magnitudes of
AXJ~1510$+$0742 and AXJ~1951$+$5609 were summarized in Table~6. The
uncertainties include not only the scatter in the conversion factors
but also the uncertainties caused by sky background determination in
the final object frame.

\section{Detailed Description on Individual Sources}

This section summarizes  the optical, near infrared, and
hard X-ray properties of each object in our sample. As already
mentioned, only narrow ($v~<~1000 $~km$^{-1}$) emission lines are
detected in the optical spectra except for AXJ~1951$+$5609.

\begin{description}
\item[AXJ~0223$+$4212 (z = 0.0435, 
filled square in Figs.~\ref{line_ratio_fig}, 
\ref{opt_x1_fig}, and \ref{z_pi_fig}):]

Narrow emission lines
of H$\beta$, H$\alpha$, [NII]$\lambda \lambda$ 6548, 6583,
and [SII]$\lambda \lambda$ 6717, 6731 are detected. [OIII]$\lambda$ 5007
is not detected.
The upper limit on the [OIII]$\lambda$ 5007 to
H$\beta$ flux ratio, the large [NII]$\lambda$ 6583 to H$\alpha$ flux
ratio, and the large [SII]$\lambda \lambda$ 6717, 6731 to H$\alpha$
flux ratio are consistent with the line ratios of LINERs
(Fig.~\ref{line_ratio_fig}). The narrow emission lines are not
resolved.
The estimated 2--10~keV luminosity is $3\times 10^{41}$~\erg\ from the
Chandra flux. This is the lowest luminosity object in this sample and
its luminosity is similar to the lowest luminosity objects found in
deep Chandra surveys and the very lowest luminosity Seyfert I galaxies. 
The optical image of the object shows two spiral
arms and the object is classified as late-type spiral (Sb) galaxy.

\item[AXJ~0431$-$0526 (z = 0.0596, filled circle):]

Narrow emission lines of  [OIII]$\lambda \lambda$ 4959, 5007,
H$\alpha$, [NII]$\lambda \lambda$ 6548, 6583, and [SII]$\lambda \lambda$ 6717,
6731 are detected. H$\beta$ emission is not detected. The lower limit
on the [OIII]$\lambda$ 5007 to H$\beta$ flux ratio, the large
[NII]$\lambda$ 6583 to H$\alpha$ flux ratio, and the large
[SII]$\lambda \lambda$ 6717, 6731 to H$\alpha$ flux ratio are
consistent with the line ratios of Seyfert IIs
(Fig.~\ref{line_ratio_fig}).  The narrow emission lines are not
resolved and with an upper limit of $<$~500~km~s$^{-1}$ are only
consistent with the narrowest line widths found in Seyfert IIs.
The optical image shows the object is an early type galaxy (E or S0). 

\item[AXJ~1025$+$4714 (z = 0.0617, filled hexagon):]

Narrow emission lines of H$\beta$, [OIII]$\lambda \lambda$
4959, 5007, H$\alpha$, [NII]$\lambda \lambda$ 6548, 6583,
and [SII]$\lambda \lambda$ 6717, 6731 are detected.
The intermediate [OIII]$\lambda$ 5007 to
H$\beta$ flux ratio, the large [NII]$\lambda$ 6583 to H$\alpha$ flux
ratio, and the large [SII]$\lambda \lambda$ 6717, 6731 to H$\alpha$
flux ratio fall in the region of Seyfert IIs but are also consistent with
LINERs within the error (Fig.~\ref{line_ratio_fig}). The narrow
emission lines are not resolved. In the blue side of the H$\alpha$
emission line, there is a hint (4$\sigma$) of a broad emission line
with velocity width of ($3100~\pm~600$~km~s$^{-1}$) (compare
[NII]$\lambda$ 6548 + H$\alpha$ profile of the object with those of
AXJ~0223+4212 and AXJ~0431$-$0526, see Fig.~\ref{finding_fig}). The
X-ray luminosity is  more than an order of magnitude larger than
that of typical LINERs ($\simlt 10^{42}$ \erg ,
\cite{Terashima2002}). This implies that the object is more like a
Seyfert II rather than a LINER. The host galaxy is E or S0.

\item[AXJ~1510$+$0742 (z = 0.4595, filled triangle):]

Narrow emission lines of [OII]$\lambda$ 3727 and
[OIII]$\lambda \lambda$ 4959, 5007  are detected.
The presence of an H$\beta$ emission line is rather marginal;
the significance of the line is about 2.5$\sigma$ due to the
dip features seen at the both side of the emission-like feature.
Thus the lower limit on [OIII]$\lambda 5007$/H$\beta$ seems to be
close to the true value.
The [OIII]$\lambda 5007$/H$\beta$  ratio implies that the
emission lines originate from star-forming regions presumably
with low-metallicity or Seyfert/LINER nucleus.
The H$\alpha$--[NII]$\lambda
\lambda$ 6548, 6583 wavelength range is not covered, and thus we could
not distinguish between either a Seyfert II/LINER or a low-metallicity
star-forming galaxy from the line ratio analysis.
The [OIII]$\lambda$ 5007 emission line is resolved, and the velocity
width is estimated to be $500~\pm~50$~km~s$^{-1}$,
 typical of the velocity width of narrow
[OIII]$\lambda$ 5007 emission lines seen in Seyfert galaxies (Whittle 1992)
and  much larger than those of dwarf star-forming galaxies.
In addition, the absolute magnitude of $\sim -22$ mag also
supports the Seyfert II/LINER interpretation.
Finally the large X-ray luminosity ($2\times10^{44}$ \erg ) suggests 
that the
object is a (high luminosity) AGN rather than a LINER. The optical
image is slightly ($< 1^{\prime\prime}$) extended.

\item[AXJ~1951$+$5609 (z = 0.910, filled diamond):]

The object show broad MgII $\lambda$ 2800 emission line with velocity
width of 5800~$\pm$~500~km~s$^{-1}$.  The object-frame equivalent
width of the broad MgII $\lambda$ 2800 line is $60^{+40}_{-20}$~\AA~
 consistent with those of optically-selected QSOs ($50~\pm~29$~\AA;
\cite{Francis1991}). The near-infrared ($J-Ks=1.24$) and
optical-to-near-infrared ($R-Ks=2.5$) colors of this object are also
consistent with the colors of optically-selected QSOs at redshift
around 1. The X-ray luminosity of $7\times10^{44}$ \erg\ is the
largest in our sample. The Chandra spectrum is apparently harder than
a canonical power law photon index of 1.7 and the presence of
absorption of $\simlt 10^{22}$ cm$^{-2}$ is suggested, although its
significance is marginal within the statistics of our data.

\end{description}

\section{Discussion}

\subsection{X-ray Time Variability}

Two out of the seven ASCA sources observed by Chandra are at least ten
times weaker than seen in the ASCA data. To check if this is
attributable to selection effects in the ASCA data and/or the result
of source confusion, we re-examined the results of Monte Carlo
simulations as performed in \S~3 of \citet{Ueda2001}. With the same
selection criteria for significance (above 4.5$\sigma$ in the 2--10
keV band and below 4.5$\sigma$ in the 0.7--2~keV band) and flux
($<3\times10^{-13}$ \ergs\ in the 2--10~keV band) as for the present
sample, we find that the probability that the flux measured with the
ASCA GIS is more than 10 times larger than the real flux to be 6\%,
assuming that 20\% of sources in the sky are highly absorbed at these
flux levels (\cite{Akiyama2000}). Hence, the expected number of false
detections in our sample is 0.4. Note that the requirement of ``no
detection in the soft band'' works to increase the fraction of false
detections in the sample if the majority of sources have non-absorbed
spectra, because unlike non-absorbed sources, the statistical
fluctuation is independent of the energy bands. Given the areal
density of Chandra sources at a flux level of 10$^{-14}$~\ergs\ of
$\sim$~200~deg$^{-2}$ the probability that an unrelated source will
fall in the $\sim$~1~arcmin$^{2}$ AMSS error circle is only 0.05. Thus
it is unlikely that these Chandra sources are completely unrelated to
the AMSS objects.

Thus, it is likely that these objects truly have a large variability
range, although we do not rule out the possibility that either of the
two is not real. Besides the two sources, AXJ~0223$+$4212 also shows a
similar amplitude of variability. While detailed variability studies
on times scales of years are only available for the brightest 25 AGNs,
it is not unusual for objects to vary by factors of five over time
scales of years and factors of ten are not unknown (e.g.,
\citet{Peterson2000} for NGC4051 and \citet{Weaver1996} for
NGC2992). Thus it is not clear if the pattern of variability of these
objects is unusual. However the large amplitude of variability confirms that
these objects are indeed AGNs.

\subsection{Summary of Identification and Comparison with Other Surveys}

Here we summarize the results of identification for the five Chandra
sources. Three show either weak or absent optical narrow emission
lines and are at very low redshifts of $<$z$>$~$\sim$~0.06
(AXJ~0223$+$4212, AXJ~0431$-$0526, AXJ~1025$+$4714), one is a narrow
line object at z~=~0.460 (AXJ~1510+0742), and the other is a broad
line object at z~=~0.910 (AXJ~1951+5609). The optical line ratios,
line widths, and the X-ray luminosity of the four narrow line objects
indicate that they are consistent with their identifications as
Seyfert II or LINER. Their X-ray spectra show strong evidence for
high column densities.

The hard X-ray flux ($f_{\rm X}$) and optical magnitudes ($f_R$) of
the identified objects are plotted in Fig.~\ref{opt_x1_fig}. In the
same figure, we plot hard X-ray selected AGNs from HEAO1-A2
(triangles), the AMSS (circles), the ALSS (squares), and Chandra Deep
Field North (asterisks and crosses, crosses means upper limit on the
optical magnitude). Note that the HELLAS sample (\cite{LaFranca2002})
falls on a similar
range to the whole AMSS sample (see \cite{Akiyama2002}). As seen from
the figure, there is a wide variety of the $f_X/f_R$ ratio in our
sample: AXJ~1510+0742 and AXJ~1951+5609 have hard X-ray to optical
flux ratio close to that of the optically-faint X-ray sources seen by
Chandra and XMM ($\log f_X/f_R \simgt +1$), while the three low
redshift sources show a much smaller ratio ($\log f_X/f_R < 0$). We
defer to a later paper a detailed discussion of the origin of the wide
range in the X-ray to optical flux ratios.

Fig.~\ref{z_pi_fig} the redshift versus apparent photon index plot of
our sample together with the ALSS sample, a flux limited complete
sample at similar flux levels (\cite{Akiyama2000}). As seen in this
figure, three sources in our Chandra sample are even harder than
the hardest X-ray source in the ALSS (AXJ~131501+314) in terms of the
best-fit photon index. For comparison, Table~\ref{lss_tab} summarizes
the optical and X-ray properties of the five ``hardest'' sources in
the above ALSS sample. These ALSS hard sources are all located at low
redshifts of $z<0.4$ and are identified as four narrow line AGNs and
one broad line AGN (AXJ~130926+2952). Equivalent widths of a narrow
emission line of H$_\alpha$ and [OIII]$\lambda5007$ are also given in
Table~\ref{lss_tab} (for line ratios see Table~3 of
\cite{Akiyama2000}). Except for AXJ~130926+2952, which shows a strong
[OIII]$\lambda5007$ line, their equivalent widths are similar to those
of the three nearby sources in our Chandra sample. Similar to the Chandra sample 
(see next section) most of the ALSS
hard sources have the morphology of elliptical or early type spiral
galaxies and no clear nucleus is visible in the optical images.

\subsection{Hard X-ray Population of ``Normal'' Galaxies}

In this subsection, we discuss the nature of the three nearby sources
in our AMSS-Chandra sample (AXJ~0223$+$4212, AXJ~0431$-$0526, and
AXJ~1025$+$4714), which are identified as  Seyfert IIs or a LINERs. As
we describe below, in many respects these sources are apparently
``normal'' galaxies, and could be considered to represent the major
population of very hard sources at fluxes of $10^{-13}$ \ergs\ (2--10
keV). They are all in the nearby universe with an average redshift of
0.06 and have a very low hard X-ray to optical flux ratio ($\log
f_X/f_R < 0$). As shown in Fig.~\ref{opt_image_fig}, these objects are
clearly resolved and show no signs of a luminous
nucleus. Interestingly, as opposed to virtually all optically selected
Seyfert IIs, 80\% of which are Sa or later in the sample of
\citet{Ho1997} and \citet{McLeod1995}, the host galaxies of
AXJ~0431$-$0526 and AXJ~1025$+$4714 have the morphology of E or
S0. This may be related to the fact that these X-ray selected sources
have larger luminosities compared with the optically selected Seyfert
IIs.

They all have similar $J-K$ colors, consistent with that of old
elliptical galaxies at z~$\sim$~0.05 ($J-K=1$), and bluer than
optically-selected QSOs at z~$\simlt$~0.2 ($J-K=1.5-2$;
\cite{Elvis1994}).  Additionally, the existence of NaD and Mgb
absorption lines in the optical continuum spectra of the objects
suggests that their continuum in the optical wavelength is dominated
by emission from host galaxy.  The X-ray to $K-$band flux ratio and
the X-ray to optical flux ratio of these objects are smaller than
normal QSOs. Even if the nuclei of AXJ~0223$+$4212, AXJ~0431$-$0526,
and AXJ~1025$+$4714 are not affected by optical extinction (have
similar $f_X/f_R$ or $f_X/f_K$ ratio to normal QSOs), the host galaxy
components of these objects are brighter than the expected optical
flux from the nucleus and may explain the relative weakness of the
optical emission lines and the absence of a blue continuum by dilution
of the nuclear component in the beam by starlight.

The optical line widths and equivalent widths themselves do not distinguish
these objects from normal (i.e., non-AGN) galaxies. The upper limits
of $\sim$~500~km~s$^{-1}$ on the line widths are consistent with the
distribution of line widths in Seyfert~II galaxies (Figure~4 of
\cite{Koski1978}) which has a average value of 600~km~s$^{-1}$ with a
250~km~s$^{-1}$ variance (see also \citet{Whittle1992} who shows that
OIII is a bit narrower with a median value of 325~km~s$^{-1}$ with a
variance of 200~km~s$^{-1}$) and thus is not definitive in the
classification of the objects. The low equivalent width of the lines
and the fact that their intensity is similar to that found in a field
galaxy sample (\cite{Tresse1999}, \cite{Carter2001}) again does not
distinguish these objects. Similarly the luminosity of the lines is
consistent with that seen in field galaxy samples. On the other hand,
as we have shown above, the line ratios of two objects lie in the
ionization range seen for many Seyfert~IIs and thus are indicative of
AGN, and one has line ratios indicative of LINER activity. 
Such objects have been already detected in optical spectroscopic
surveys of field galaxies: \citet{Carter2001} find $\sim$~15~\% of all
field galaxies have "AGN-like" emission line ratios.

The colors of the three nearby objects are completely consistent with
that of stars with no sign of a non-thermal continuum. However, these
galaxies are optically luminous with Mv~$\sim$~$-$(21--22) and the
relatively low luminosity active nuclei in these objects may be hidden
by the glare of the starlight despite the low redshifts. Using the
\citet{Elvis1994} spectral energy distribution one predicts that the
effective optical magnitude of the active nucleus would be
$\sim$~17--19~mag assuming that the nucleus is not absorbed and thus
rather difficult to detect from ground based observations. The fact
that these rather bright nearby galaxies harbor ``optically
invisible'' AGN shows the incompleteness of optical active galaxy
searches. Since the colors of these objects also show no abnormalities
they would be missed by color selected surveys like the Sloan digital
sky survey. It is also possible that the nuclei are absorbed in the
optical and UV bands by the same matter that absorbs the X-ray
continuum, reducing the observed optical flux. The effective reddening
corresponding to the X-ray column densities of
(0.5--15)~$\times$~10$^{22}$~cm$^{-2}$ is Av~$\sim$~ 2--70 mags thus
effectively eliminating the optical-UV flux.

\subsection{Contribution to the X-ray background}

Our results indicate that ``very hard'' X-ray sources at fluxes of
$\sim~10^{-13}$~\ergs\ (2--10~keV) are mostly nearby $z < 0.5$ sources
with a column density of $\sim~10^{22-23}$~cm$^{-2}$, rather than high
redshift (hence high luminosity) objects with a higher column density
of $\sim~10^{23-24}$~cm$^{-2}$. This fully supports the result of the
ALSS and is a strong constraint on  the unified models of the 
X-ray background. According to the
calculation of \citet{Comastri2001} and \citet{Gilli2001}, about half of
sources with a column density $>$~10$^{22}$~cm$^{-2}$ should have redshifts
greater than 1 at a flux of $\sim~1~\times$~10$^{-13}$~\ergs . We have
to remember, however, that there could be bias against high redshift
objects having moderate column densities in our Chandra sample because
our selection criteria were based on the ``apparent'' hardness: due to
K-correction absorbed sources at higher redshifts show softer spectra
in the observer frame. Precise measurement of the column densities of
slightly less hard sources detected in the ALSS and AMSS
(\cite{Akiyama2002}) is important for a definitive test of the unified
model.

The optical counterparts of the hard, faint ASCA sources discussed in
this paper have rather different properties from those of soft X-ray
and optically selected AGN in the sense that they have earlier
morphological types, do not possess a luminous optical or IR nucleus,
have near-infrared colors similar to that of normal galaxies, and show
only weak narrow emission lines. These unusual properties have already
seen in the ALSS hard sample (Table~\ref{lss_tab}) and in the HELLAS
survey at a similar X-ray flux level (\cite{Maiolino2000}) as well as
in the deeper Chandra fields (e.g., \cite{Barger2001}).

\section{Summary}

We present the results of the first third of our Chandra follow-up of
faint hard X-ray selected ASCA sources. We have optical counterparts
for five of the seven objects, since two of them were too faint to
uniquely associate the Chandra with the ASCA source. Of these five
optical counterparts three of them show either weak or absent optical
lines of low physical and equivalent width and are at low redshift
$<$z$>$~$\sim$~0.06. One of them is a z~=~0.460 object with only
narrow lines and one is a broad line object at z~=~0.910.  All the
narrow line objects show strong evidence for absorption in their X-ray
spectra, and their line ratios are consistent with a Seyfert II/LINER
identification as are the line widths. The three nearby objects have
the colors of normal galaxies and apparently the light is dominated by
stars. This could be due to the extinction of the underlying nuclear
continuum by the same matter that absorbs the X-rays and/or due to the
dilution of the central source by starlight. These results suggest
that X-ray sources that appear as ``normal'' galaxies in optical and
near-IR bands with weak (but AGN-like) narrow emission lines
significantly contribute to the hard X-ray background, supporting the
previous results of the ALSS and the HELLAS. This population of
objects has a high space density and probably dominates the entire
population of active galaxies.

\newpage

\begin{table}
\caption{Target list of our observations with Chandra. The corresponding source
name in the AMSS catalog (\cite{Ueda2001}), if present, is given in 
the parenthesis. The position and flux are determined with ASCA.
The ASCA error radius is 1--1.5 arcmin.}
\begin{center}
\begin{tabular}{lccccc}
\hline
Target(Cataloged Name)          & RA, DEC   & \multicolumn{2}{c}{ASCA flux$^{*}$} & \multicolumn{2}{c}{Chandra Observation}\\
                                & (J2000)   & 2--10~keV  & 0.7--2~keV & Exposure& Date\\ \hline
AXJ 0223+4212(1AXG~J022353+4212)& 35.9719, 42.2083   & 3.0 & 0.21 & 6.48 ks &05 Sep. 2000\\ 
AXJ 0431$-$0526                 & 67.9280, $-$5.4365 & 2.7 & $<$0.06$^{\dagger}$ & 5.97 ks &03 Oct. 2000\\ 
AXJ 1025+4714(1AXG~J102557+4713)&156.4883, 47.2316   & 2.2 & $<$0.08$^{\dagger}$   & 4.77 ks &07 Jun. 2000\\ 
AXJ 1227+4421                   &186.9139, 44.3541   & 1.5 & 0.25                 & 7.31 ks &04 Oct. 2000\\ 
AXJ 1510+0742                   &227.5974, 7.7006    & 2.5 & $<$0.37$^{\dagger}$   & 6.46 ks &01 Jun. 2000\\ 
AXJ 1951+5609(1AXG~J195105+5610)&297.7717, 56.1677   & 2.0 & 0.37 & 8.47 ks &25 Apr. 2000\\ 
AXJ 2018+1139(1AXG~J201822+1139)&304.5932, 11.6562   & 1.7 & 0.15 & 8.52 ks &30 Aug. 2000\\ \hline
\end{tabular}
\end{center}
$^{*}$: in units of 10$^{-13}$ erg cm$^{-2}$ s$^{-1}$. The flux is calculated from the best-fit count 
rate in the same band assuming a photon index of 1.7. 
Note that all the detection significance in the soft band (0.7--2~keV) is below 4.5~$\sigma$. \\
$^{\dagger}$: 1$\sigma$ upper limit.
\label{list_tab}
\end{table}

\begin{table}
\caption{X-ray data of five Chandra sources that are clearly
associated with the ASCA sources. Photon Indices are calculated from
the hardness ratio between the 2--10~keV and 0.7--2~keV counts,
correcting for Galactic absorption. The column density (at the source
redshift) is determined assuming a photon index of 1.7. The flux is an
observed one (i.e., not corrected for absorption nor for redshift),
while the luminosity is corrected for absorption and is given in the
2--10 keV source frame, both calculated from the best-fit absorbed
power-law model with a photon index of 1.7. }
\label{xray_property_tab}
\begin{tabular}{cccccccc}
\hline
Target      & RA, DEC  & Cts/ks         & Cts/ks         & Photon
& N$_{\rm H}^{\dagger}$ & Flux  & Luminosity \\
            & (J2000)   & 0.7--2~keV       & 2--10~keV        & Index$^{*}$
&  [10$^{22}$~cm$^{-2}$]     & [erg cm$^{-2}$ s$^{-1}$] & [erg s$^{-1}$] \\
\hline
AXJ 0223+4212 & 35.9737, 42.2061 & 0.46 & 0.77 &  
0.3$\pm$0.7 & 3.1$^{+3.6}_{-2.1}$    & $4.0 \times 10^{-14}$ &
$2.5 \times 10^{~41}$    \\
AXJ 0431$-$0526 & 67.9134, $-$5.4320 & 0.34 & 8.9 &  
$-$1.2$\pm$0.1 & 14.4$^{+6.1}_{-5.1}$ & $7.1 \times 10^{-13}$ &
$2.2 \times 10^{~43}$ \\
AXJ 1025+4714 & 156.4774, 47.2399 & 0.0              & 8.8&
$<-$1.2     & 9.3$^{+4.0}_{-2.5}$ & $6.7 \times 10^{-13}$ & $1.9 \times
10^{~43}$ \\
AXJ 1510+0742 & 227.6089, 7.6899 & 0.15 & 2.3 & 
$-$1.1$\pm$0.3 & 13.9$^{+10.6}_{-6.4}$ & $1.0 \times 10^{-13}$ &
$1.8 \times 10^{~44}$\\
AXJ 1951+5609 & 297.7547, 56.1647 & 7.0 & 3.9 &
1.4$\pm$0.2 & 0.63$^{+1.09}_{-0.63}$ & $1.2 \times 10^{-13}$ &
$7.2 \times 10^{~44}$ \\ \hline
\multicolumn{8}{l}{$^{*}$ : The errors correspond to 1$\sigma$.} \\
\multicolumn{8}{l}{$^{\dagger}$: A photon index of 1.7 is assumed. The errors correspond to 90\% confidence level.} \\
\end{tabular}
\end{table}

\begin{table}
\caption{Journal of optical spectroscopic observations.}
\begin{center}
\begin{tabular}{clll}
\hline
Name            & Telescope & Date     & Integration time \\ \hline
AXJ~0223$+$4212 & UH88      & 05 Oct. 2000 & 300~s \\
AXJ~0431$-$0526 & UH88      & 06 Oct. 2000 & 600~s \\
AXJ~1025$+$4714 & UH88      & 20 Mar. 2001 & 900~s  \\
AXJ~1510$+$0742 & UH88      & 21 Mar. 2001 & 900~s $\times$ 3 \\
AXJ~1951$+$5609 & KPNO2.1m  & 20 Sep. 2000 & 900~s+1800~s $\times$ 2 \\
\hline
\end{tabular}
\end{center}
\label{optjarnal_tab}
\end{table}

\begin{table}
\caption{Basic parameters of the sample.\label{basic}}
\begin{center}
\begin{tabular}{llllll}
\hline 
Name          & Redshift       & $m_{R}$ & $M_{V}^{*}$ & Note \\ \hline
AXJ~0223$+$4212 & $0.0435~\pm~0.0001$ & 16.0 & $-21.1$ & Sb \\
AXJ~0431$-$0526 & $0.0596~\pm~0.0002$ & 15.6 & $-22.2$ & E/S0 \\
AXJ~1025$+$4714 & $0.0617~\pm~0.0001$ & 16.9 & $-20.9$ & E/S0 \\
AXJ~1510$+$0742 & $0.4595~\pm~0.0001$ & 20.1 & $-22.1$ & Extended \\
AXJ~1951$+$5609 & $0.9096~\pm~0.0009$ & 19.5 & $-24.2$ & \ldots \\
\hline
\multicolumn{6}{l}{$^{*}$: $V-R$=0.22 mag (power law index of --0.5) is assumed. }\\
\end{tabular}
\end{center}
\label{basic_tab}
\end{table}

\begin{table}
\caption{Equivalent widths of narrow emission lines in units of $\AA$ and their line-intensity ratios. [NII], [SII], and [OIII] denotes the [NII]$\lambda$6583, [SII]$\lambda\lambda$6716+6731, and [OIII]$\lambda$5007 lines, respectively.
} 
\label{line_equi}
\begin{center}
\begin{tabular}{ccccccccc}
\hline
Name  & H$_{\alpha}$ & [NII] & [SII] & [OIII] & H$_{\beta}$ &
$\log \frac{\rm [NII]}{\rm H_{\alpha}}$ & 
$\log \frac{\rm [SII]}{\rm H_{\alpha}}$ & 
$\log \frac{\rm [OIII]}{\rm H_{\beta}}$ \\ \hline
AXJ~0223$+$4212 & $20\pm1$ & $20\pm1$ & $11\pm2$ & $<5.7$ & $5.9\pm2.0$ &
$0.00^{+0.04}_{-0.04}$ & $-0.26^{+0.10}_{-0.11}$ & $<0.16$ \\
AXJ~0431$-$0526 & $7.3\pm0.4$ & $7.4\pm0.4$ & $5.6\pm0.6$ & $5.8\pm0.8$ & $<0.8$ &
$0.00^{+0.05}_{-0.04}$ & $-0.12^{+0.07}_{-0.07}$ & $>0.80$ \\
AXJ~1025$+$4714 & $12\pm1$ & $5.8\pm1.0$ & $7.7\pm1.1$ & $8.4\pm0.9$ & $2.3\pm0.8$ &
$-0.31^{+0.10}_{-0.12}$ & $-0.19^{+0.09}_{-0.10}$ & $0.56^{+0.23}_{-0.18}$ \\
AXJ~1510$+$0742 & no data & no data & no data & $69\pm3$ & $<23$ &
\ldots & \ldots & $>0.45$ \\
\hline
\end{tabular}
\end{center}
\end{table}

\begin{table}
\caption{Near-infrared magnitudes of our sample. The data of the first
3 sources are from 2MASS catalog, and the rest from our own
observations. The number in parenthesis indicates a 1$\sigma$ error.}
\begin{center}
\begin{tabular}{lllll}
\hline 
Name           & $J$    & $H$    & $K_s$  & $J-K_s$ \\ \hline
AXJ~0223$+$4212 & 14.53 (0.05) & 13.74 (0.06) & 13.35 (0.06) & 1.18 (0.08) \\
AXJ~0431$-$0526 & 14.52 (0.04) & 13.83 (0.05) & 13.36 (0.05) & 1.16 (0.06) \\
AXJ~1025$+$4714 & 15.12 (0.06) & 14.49 (0.07) & 13.94 (0.08) & 1.18 (0.10) \\
AXJ~1510$+$0742 & 18.05 (0.05) & 17.29 (0.07) & 16.72 (0.05) & 1.33 (0.07) \\
AXJ~1951$+$5609 & 18.22 (0.05) & 18.03 (0.07) & 16.98 (0.05) & 1.24 (0.07) \\
\hline
\end{tabular}
\end{center}
\label{2mass_tab}
\end{table}

\begin{scriptsize}
\begin{table}
\caption{Summary of optical/X-ray properties of hard sources in the ALSS.}
\label{lss_tab}
\begin{center}
\begin{tabular}{cccccccl}
\hline 
Name           & z & $m_R$ & $f_{\rm X}$ (2--10 keV)& log($f_{\rm X}/f_R)$  & 
\multicolumn{2}{c}{E.W. [$\AA$]} & Morphology \\ 
	       &   &       & [ \ergs ]  &                       &
 H$_{\alpha}$ &  [OIII]$\lambda$5007 & \\\hline
AXJ~131551+3237 &0.128 &18.1 &2.6$\times10^{-13}$ &0.16 &25$\pm$3 & 37$\pm$3 & E/S0\\
AXJ~131501+3141 &0.072 &15.6 &4.8$\times10^{-13}$ &$-0.57$ &11$\pm$1 & 7$\pm$1 &Early-type Spiral\\
AXJ~130926+2952 &0.375 &19.9 &1.5$\times10^{-13}$ &0.64 & no data & 109$\pm$22 &E/S0 or Early-type Spiral\\
AXJ~131210+3048 &0.189 &18.4 &1.8$\times10^{-13}$ &0.11 &13$\pm$2 &16$\pm$2 & E\\
AXJ~130840+2955 &0.164 &17.3 &1.4$\times10^{-13}$ &$-0.42$ &no data$^{*}$ &8.6$\pm$0.9 & Spiral (Interacting?) \\
\hline
\multicolumn{8}{l}{$^{*}$ : Not available due to atmospheric absorption.}\\
\end{tabular}
\end{center}
\end{table}
\end{scriptsize}

\begin{figure}
\begin{center}
\FigureFile(40mm,40mm){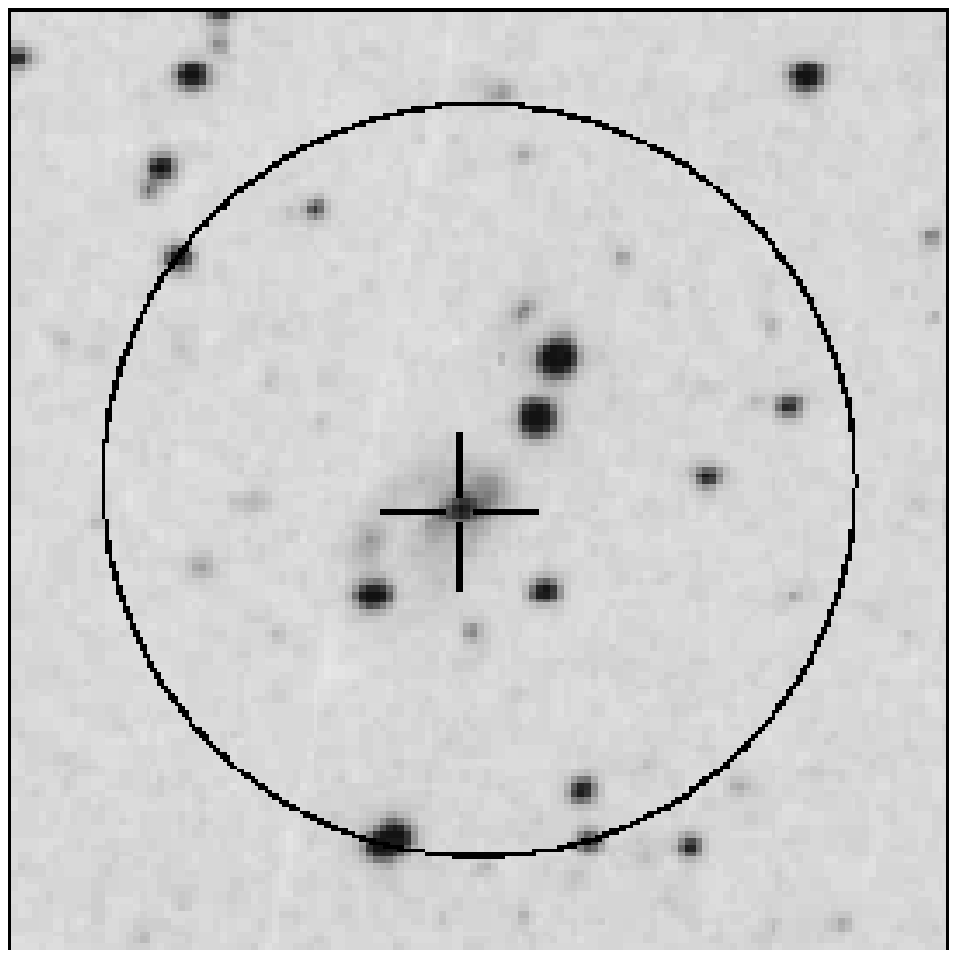}
\FigureFile(100mm,40mm){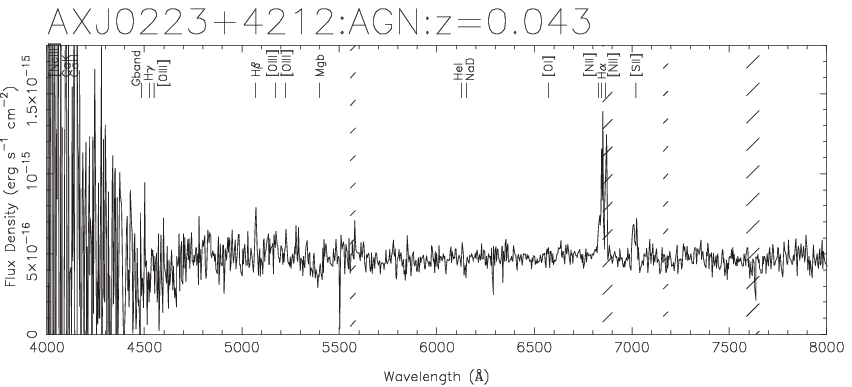}

\FigureFile(40mm,40mm){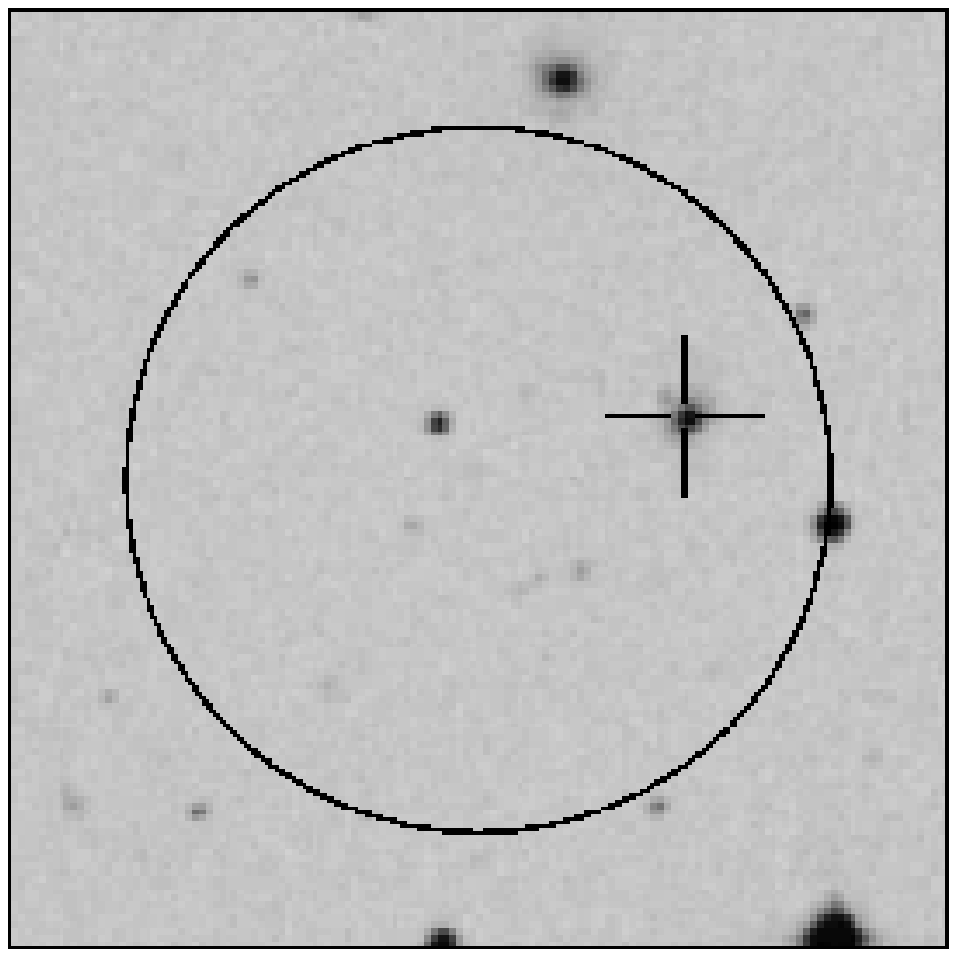}
\FigureFile(100mm,40mm){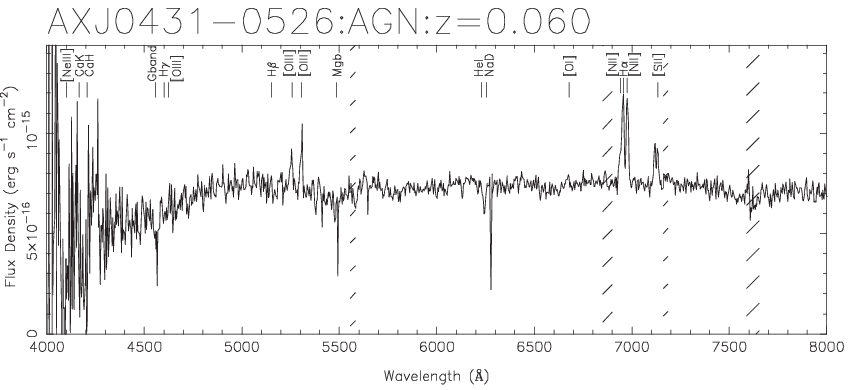}

\FigureFile(40mm,40mm){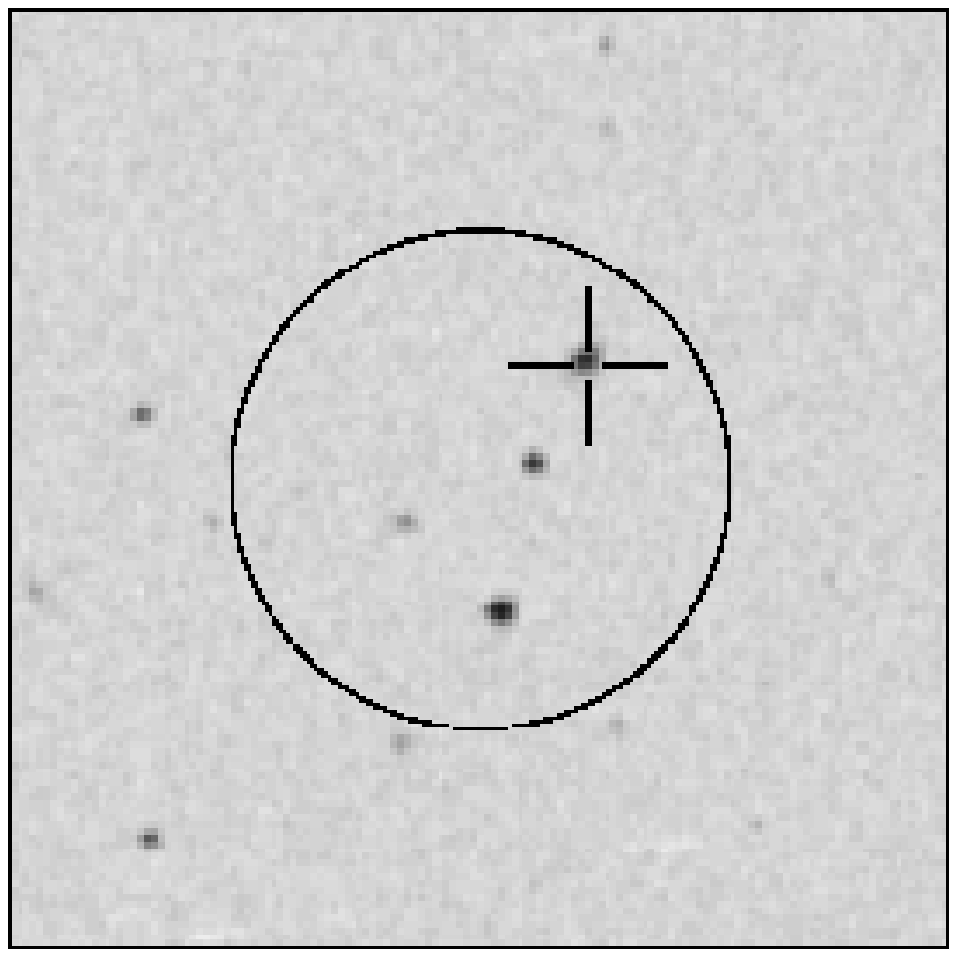}
\FigureFile(100mm,40mm){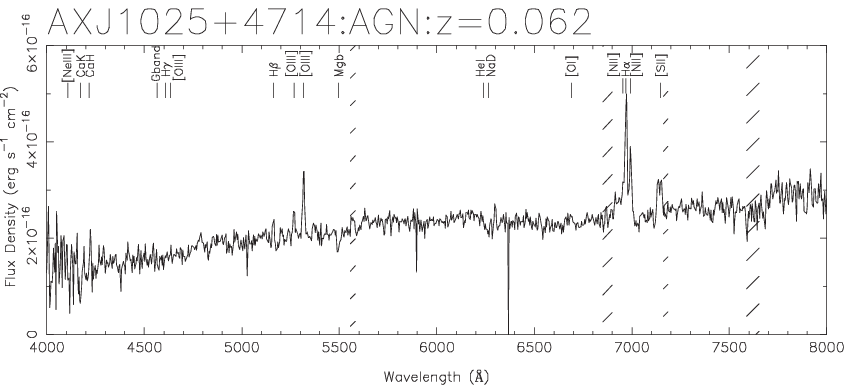}

\FigureFile(40mm,40mm){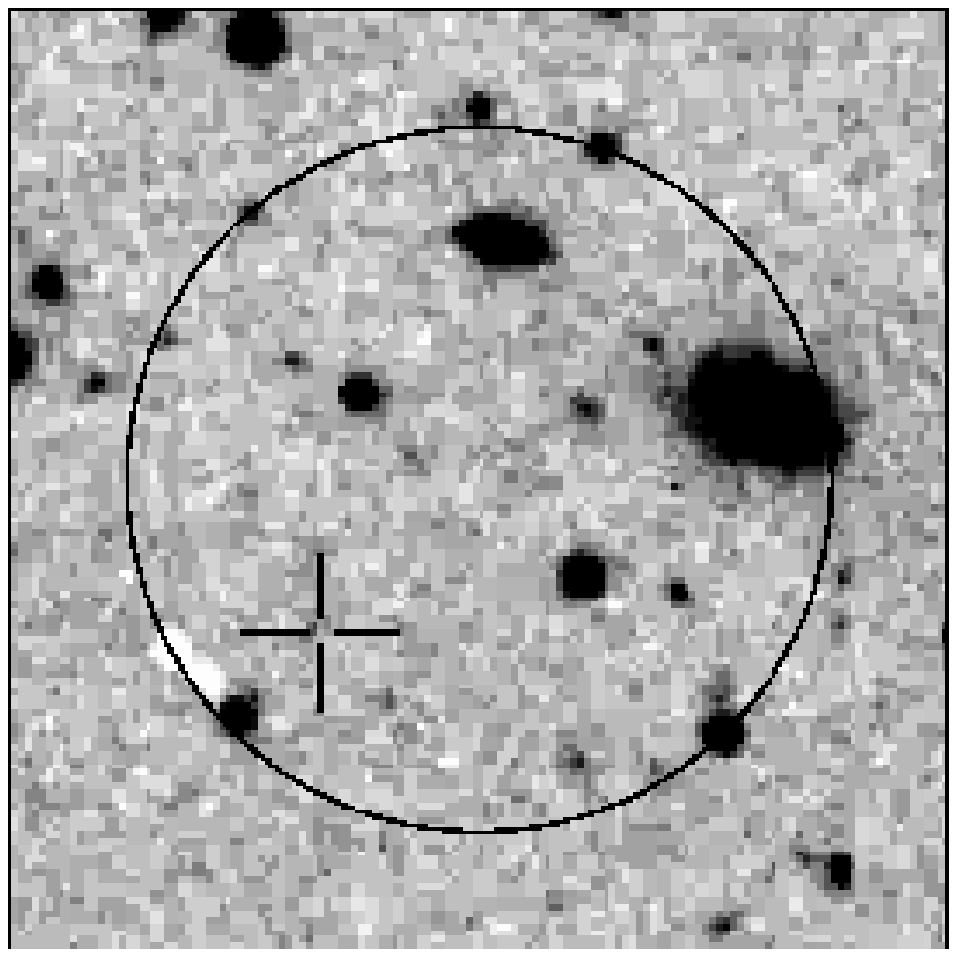}
\FigureFile(100mm,40mm){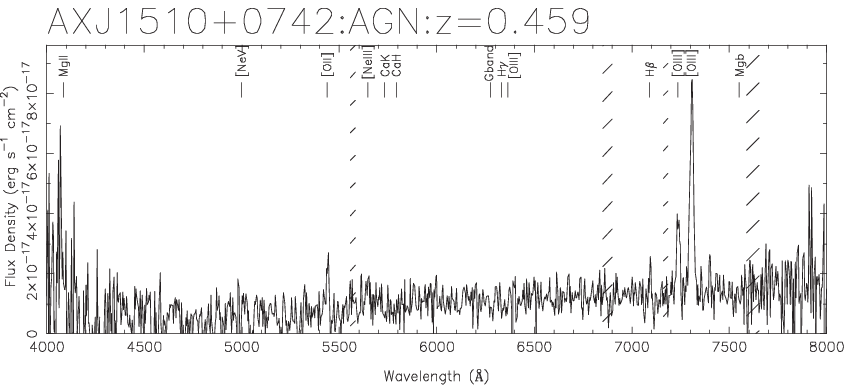}

\end{center}
\end{figure}

\begin{figure}
\begin{center}
\FigureFile(40mm,40mm){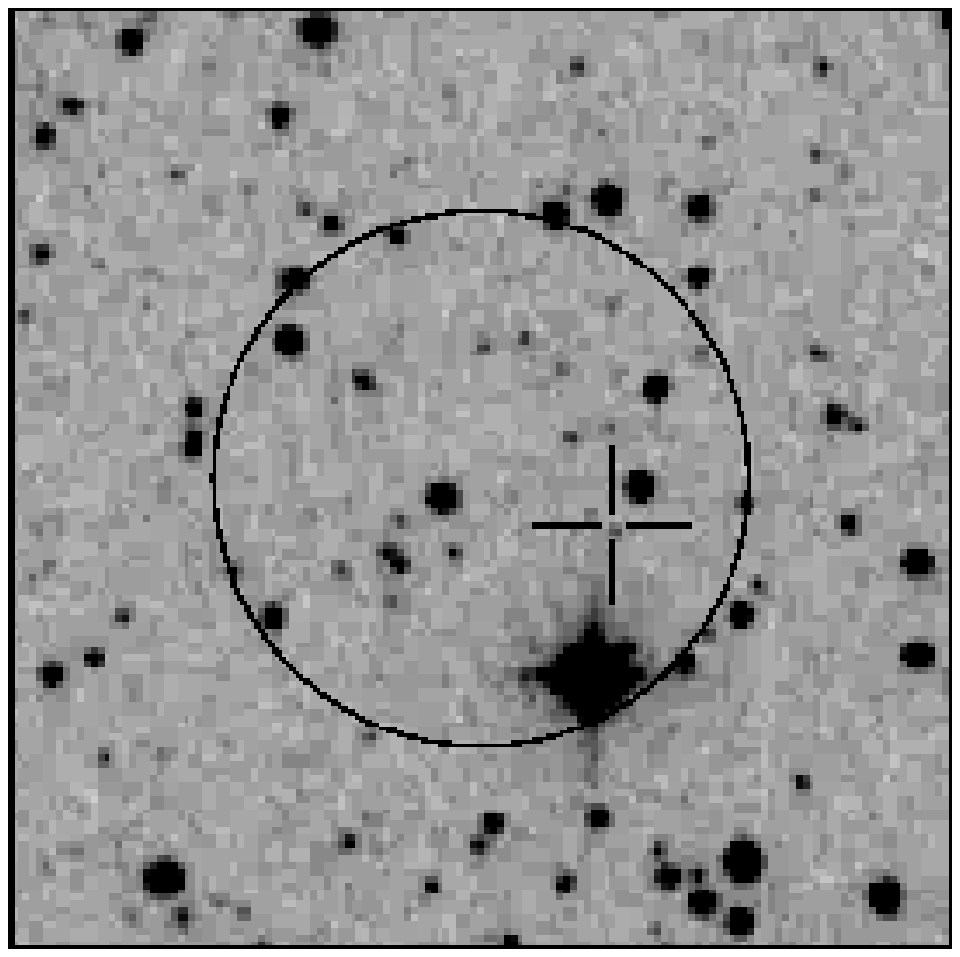}
\FigureFile(100mm,40mm){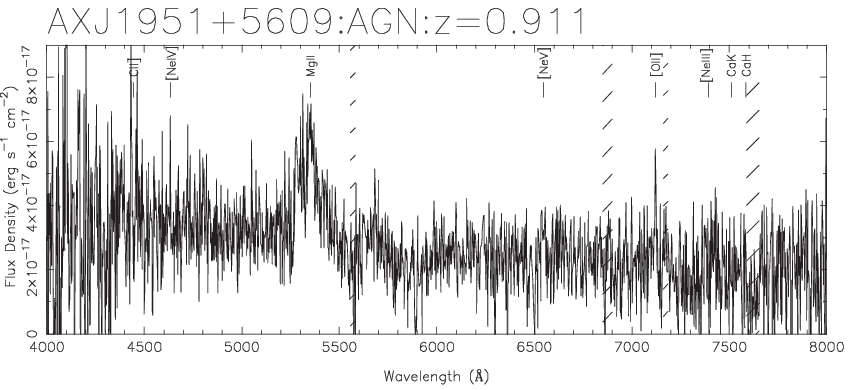}

\caption{Optical finding charts and spectra of identified objects.
The finding charts are made from Digitized Palomar Observatory Sky
Survey data.  The field of view of the finding charts is
4~\arcmin~$\times$~4~\arcmin.  The large circle corresponds to an
error circle of ASCA, and the central position of the cross
corresponds to the position of the Chandra source (with uncertainties
of $1^{\prime\prime}$). In the spectra, the positions of principal
lines are shown.  The diagonal lines show the regions which are
affected by absorption lines and emission lines of the atmosphere.}
\label{finding_fig}
\end{center}
\end{figure}

\begin{figure}
\begin{center}

\FigureFile(170mm,90mm){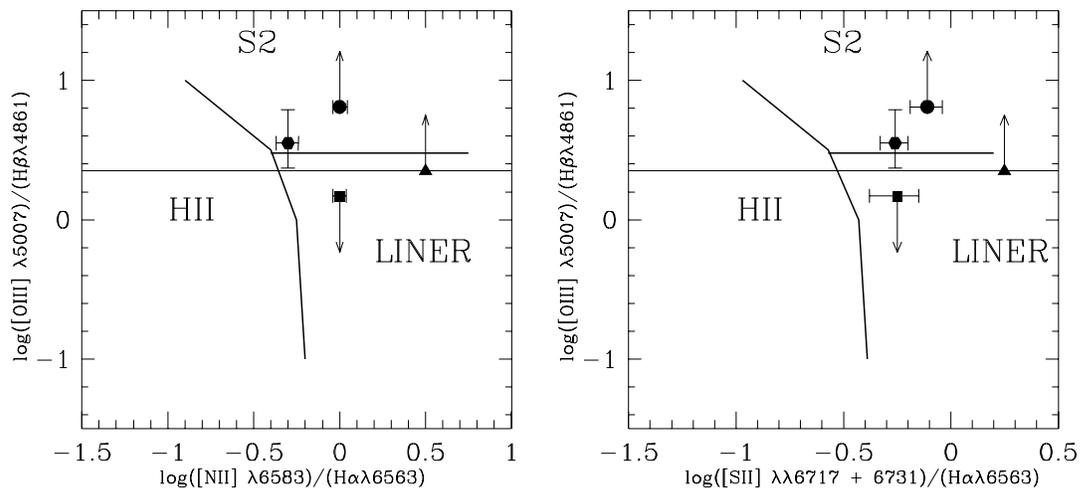}

\caption{The plot of intensity ratios of the optical narrow emission
lines (filled square: AXJ~0223+4212, filled circle: AXJ~0431$-$0526,
filled hexagon: AXJ~1025+4714, filled triangle: AXJ~1510+0742). Solid
lines in the figure represent boundaries between Seyfert II galaxies,
LINERs, and H~II region-like galaxies, taken from Veilleux \&
Osterbrock (1987) and Veilleux (1995). }
\label{line_ratio_fig}
\end{center}
\end{figure}

\begin{figure}
\begin{center}

\FigureFile(120mm,120mm){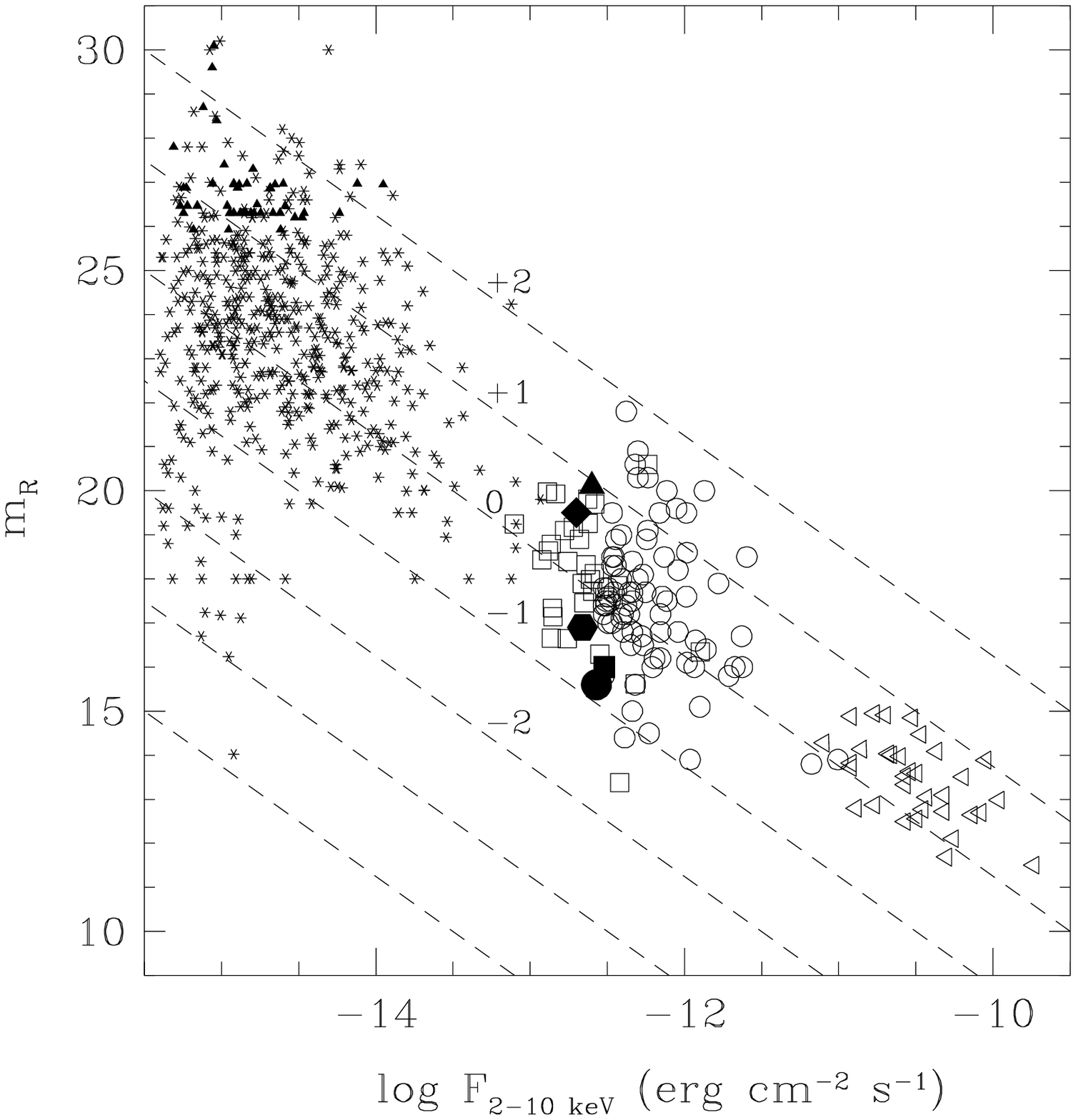}

\caption{Correlation between X-ray flux and optical magnitude 
(filled square: AXJ~0223+4212, filled circle: AXJ~0431$-$0526, filled
hexagon: AXJ~1025+4714, filled triangle: AXJ~1510+0742, filled
diamond: AXJ~1951+5609). Open triangles, open circles, and open
squares are the AGNs from HEAO1 A2 (\cite{Piccinotti1982}), 
the AMSS (\cite{Akiyama2002}), the ALSS (\cite{Akiyama2000}),
respectively. Asterisks and crosses are the source from Chandra Deep
Field North (\cite{Brandt2001}, \cite{Barger2002}), 
with crosses being upper limits on the optical magnitude
Dashed lines represent the X-ray to optical flux
ratio of $\log f_X / f_R =$ $+2$, $+1$, $0$, $-1$, $-2$, $-3$, 
and $-4$ from top to bottom.
} \label{opt_x1_fig}

\end{center}
\end{figure}

\begin{figure}
\begin{center}

\vspace{-10cm}
\FigureFile(200mm,200mm){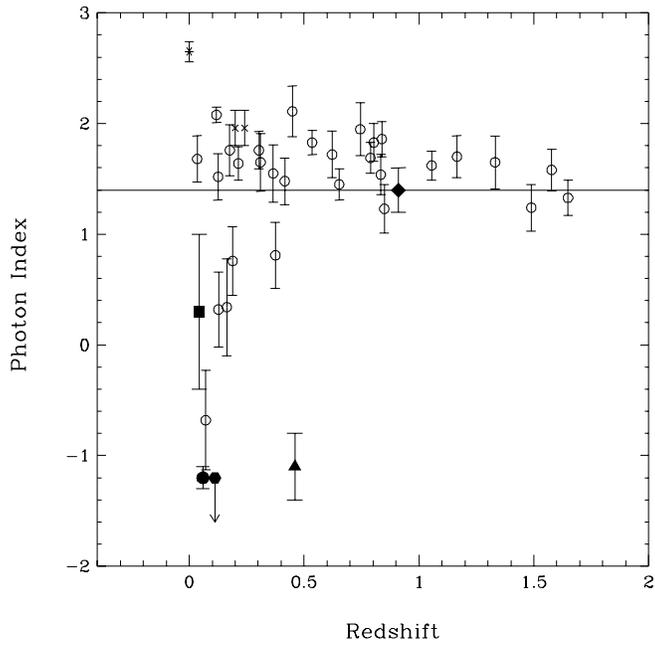}

\caption{The redshift versus an apparent photon index plot of our
sample, (filled square: AXJ~0223+4212, filled circle: AXJ~0431$-$0526,
filled hexagon: AXJ~1025+4714, filled triangle: AXJ~1510+0742, filled
diamond: AXJ~1951+5609), compared with the ALSS sample (open circles:
AGN, after Fig.~5 of \citet{Akiyama2000}).}
\label{z_pi_fig}

\end{center}
\end{figure}

\end{document}